\documentclass[a4paper]{spie}  

\usepackage{amsmath,amsfonts,amssymb}
\usepackage{graphicx}
\usepackage[colorlinks=true, allcolors=blue]{hyperref}

\title{New methods for ALMA angular-scale based observation scheduling, quality assessment, and beam shaping}

\author[a]{Dirk Petry}
\author[a]{Mar\'ia D\'iaz Trigo}
\author[b,c]{R\"{u}diger Kneissl}
\author[c]{Ignacio Toledo}
\author[a]{Stefano Facchini}
\affil[a]{European Southern Observatory, Karl-Schwarzschild-Str. 2, 85748 Garching, Germany}
\affil[b]{European Southern Observatory, Alonso de Cordova 3107, Vitacura, Santiago, Chile}
\affil[c]{Joint ALMA Observatory, Alonso de Cordova 3107, Vitacura, Santiago, Chile}

\authorinfo{Further author information: (Send correspondence to Dirk Petry)\\D. Petry: E-mail: dpetry@eso.org}

\begin{document} 
\maketitle

\begin{abstract}
Up to now, the completion of an ALMA interferometric observation is determined based
on the achievement of a given shape and size of the synthesized beam and the noise RMS in the representative spectral range. This approach with respect to the angular resolution investigates mainly the longest baselines of the interferometer and says little about the sensitivity at larger angular scales. We are exploring the ideas of angular-scale-based scheduling and quality assessment, and of angular-scale-based visibility weighting as a
step towards optimising both observation efficiency and image fidelity. 
This approach carries the imaging quality assurance into the visibility space, where interferometers record the data, and therefore simplifies many aspects of the procedure. Similarly during scheduling such detailed assessment of the expected imaging properties helps optimising the scheduling process. 
The methodology is applicable to all radio interferometers with more than ca. 10 antennas. 
\end{abstract}


\section{INTRODUCTION}
\label{sec:intro}

The Atacama Large Millimeter/submillimeter Array (ALMA)
is one of the largest astronomical projects today. In operation since end of 2011 and continuously being updated and extended, ALMA is now in observation Cycle 7. The observatory presently consists of an array of fifty 12-m antennas and an additional compact array (ACA) of twelve 7-m and four 12-m "total power" (TP) antennas to enhance ALMA's ability to image extended objects. Baseline lengths range from 7 m (minimum possible length for the ACA) up to ca. 16~km for the 12-m array.

The ALMA project is an international collaboration between Europe, East Asia, and North America in cooperation with the Republic of Chile.
The official project website for scientists is the {\it ALMA Science Portal}
{\tt http://www.almascience.org}. A detailed description of ALMA and many of the standard procedures mentioned here can be found in the Cycle 7 ALMA Technical Handbook \cite{remijan2019}.

Scientists proposing for using ALMA do not request a particular amount of observing time but define {\it science goals} which consist of sets of observation parameters to be achieved by ALMA. The main parameters are
\begin{enumerate}
    \item The coordinates of the astronomical target(s) to be observed (one of them is used as the representative target if there is more than one) or the raster pointings (in case of mosaics) and the spectral range and spectral resolution to use.
    \item The sensitivity of the observation expressed as the noise RMS to be achieved in a given frequency bandwidth centered on a given representative frequency.
    \item The angular resolution (AR) of the final image expressed as the size (diameter) of the synthesized beam of the interferometer to be achieved at the representative frequency.
    \item (for an extended, i.e. resolved target) the largest angular scale (LAS) for which the interferometer should still be sensitive. 
\end{enumerate}
If the proposal is accepted, these science goals are
translated into individual {\it scheduling blocks} (SBs). An SB is a description of an observation scan sequence complete with the necessary plan of calibrator observations. The typical observation time foreseen in an SB is 20 to 90 minutes (including all overhead for calibrations). The execution of an SB results in a so-called execution block (EB), the smallest unit of processable data for ALMA. In order to achieve a high sensitivity or meet time constraints, an SB may have to be executed more than once which results in several EBs. The set of all EBs of an SB then forms the final dataset for a given science goal, the so-called Member Observation Unit Set (MOUS). 

All EBs undergo a quality assurance (QA) procedure immediately after observation. This is called level zero QA or QA0. Rejected EBs are discarded and reobserved if possible. When all necessary EBs of an MOUS have been observed or no further observations for the given SB are possible, the MOUS is sent for level 2 QA or QA2. This final step of QA includes a complete, science-ready calibration of the data and the generation of image products for the ALMA archive. If the MOUS does not pass QA2, it is sent back for reobservation if that is still possible. Otherwise, it is delivered to the principle investigator (PI) of the proposal. More details of the QA process can be found in \cite{remijan2019,nakos2020} and references therein.

Important in the context of this paper is that the decision on QA2-pass or -fail is based on measuring whether the science goal 
parameters have been achieved. Here, the procedure adopted so far did properly measure the noise RMS and the AR, but the LAS was only assessed indirectly if at all. It was assumed that the general design of the ALMA array configurations would ensure the fulfillment of the LAS condition. The same is true for all intermediate angular scales. 

In cases of a particularly large range between AR and LAS, the SB design 
foresees a splitting of the observation into up to four SBs: up to three for the larger angular scales (using a more compact 12-m array configuration with shorter baselines and/or the ACA and the TP antennas) and one for the smaller angular scales down to the AR (using a more extended array configuration). The MOUSs resulting from the up to four SBs would then be combined offline after QA2 in the science analysis by the PI. Such a set of MOUSs observing the same target at
different spatial scale ranges is called a Group Observation Unit Set (GOUS).

In this paper we present the intermediate results of an ongoing ESO ALMA internal development study on improving our methods of assuring the achievement of the science goal parameters both in observation scheduling and in subsequent QA and final imaging.
Having realised that determining the synthesized beam size essentially only measures the achieved sensitivity for the longest baselines, we explore a more complete approach where we separately measure the achieved sensitivity in all ranges of baseline lengths, i.e. all observed angular scales, and then compare these to an expectation which we derive from the AR and LAS specified in the science goal.
We propose to perform this comparison between observation and expectation not only at the end of the observation procedure in the QA stage but already during the scheduling of multi-EB MOUSs,
dynamically choosing the array configuration for the next execution based on the baseline length distribution achieved so far.

\section{THE BASELINE LENGTH DISTRIBUTION - EXPECTATION AND OBSERVATION}
\label{sec:bldist}

One of the main concepts we need to introduce in our new approach is the Baseline Length Distribution (BLD). This is a histogram of the projected length (in meters) of the baselines formed by the antennas of the interferometer array. For each integration  (visibility) recorded in the interferometric dataset (after discarding invalid data, so-called "flagging"), one entry is made in the BLD bin corresponding to the length of the baseline used for that integration. We distinguish between 
\begin{description}
\item[unweighted baseline length distribution] - the histogram entry for one integration is given a weight equal to the integration time. 
\item[absolutely weighted baseline length distribution] - the histogram entry for one integration is given a weight equal to
the inverse per-channel noise squared (which is what is recorded in the WEIGHT column of a MeasurementSet). The WEIGHT value is proportional to the integration time, the channel bandwidth and inversely proportional to the system temperature.
\item[relatively weighted baseline length distribution] - here the histogram entry is given a weight equal to the WEIGHT value as above but normalised to the average WEIGHT of all baselines.
\end{description}
The y-axis of the histogram has units of "visibility hours" in all three cases. Commonly, the "sensitivity" of an observation is regarded as inversely proportional to the achieved RMS image noise. The sensitivity of an interferometric observation is thus proportional to the square root of the product of observation time and number of baselines used, and the value of the individual histogram column of the BLD is proportional to the square of the sensitivity achieved for that baseline length range. 

\begin{figure} [htb]
\begin{center}
\begin{tabular}{c} 
\includegraphics[height=12cm]{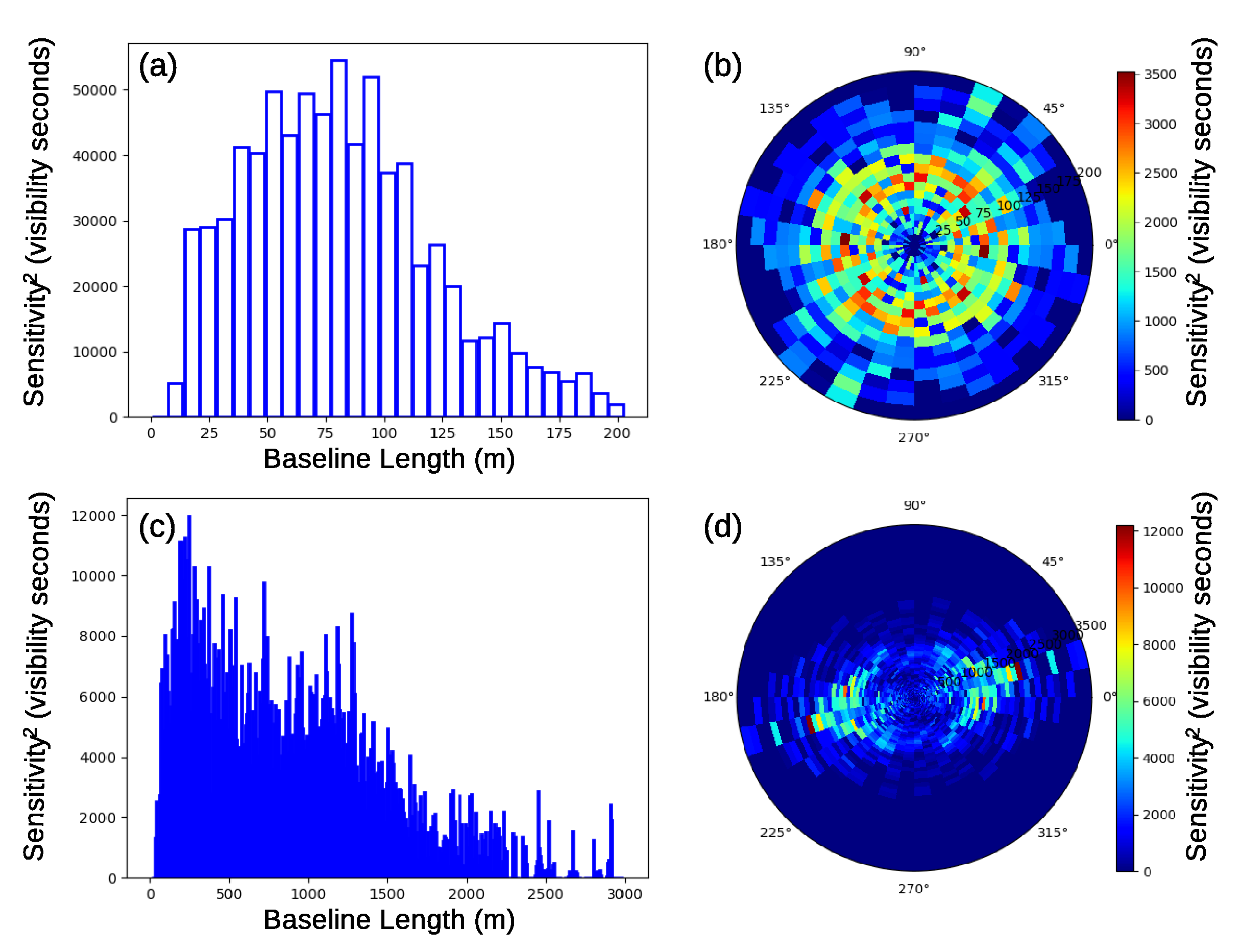}
\end{tabular}
\end{center}
\caption[]
{ \label{fig:bldistexamples} Examples of 1D and 2D baseline length distributions (with relative weighting) for (top: (a) and (b)) an ALMA 8 minute observation using 44 antennas in a compact configuration and (bottom: (c) and (d)) a  39 minute exposure using 47 antennas in an extended (hybrid) configuration which is unusually elongated. Histograms (a) and (c) both use 7 m bin width while the 2D histograms (b) and (d) use a variable radial bin width linearly growing with baseline length. See text.}
\end{figure}

Each baseline length range corresponds to a range of angular scales in the interferometric image derived from the dataset via the equation $a = \lambda / d$ where $a$ is the angular scale, $\lambda$ is the observing wavelength, and $d$ is the baseline length. The BLD is thus equivalent to a sensitivity vs. angular scale plot. Histogramming the baseline length rather than the angular scale itself, however, is more convenient in the scheduling context because the BLD of a given array configuration does not depend on the observing frequency. 
Figure \ref{fig:bldistexamples} shows examples of 1D and 2D BLDs.

For both forms of representing the sensitivity to different angular scales, there is a problem of low number of counts at the upper end of the histogram (see Fig.~\ref{fig:bldistexamples}c) and a problem of resolution at the lower end. Also logarithmically increasing the bin width does not result in a flat distribution and is less intuitive to read. Furthermore the inclusion of the important "zero spacings" from single dish observations in the first bin is awkward in a log scale. We have therefore chosen to exclusively work with the BLD representation and remedy the histogramming problems by introducing a special bin width scheme adapted to the ALMA arrays: The first two bins are fixed to 7 m width. Starting with the third bin, the bin width is increasing linearly such that it reaches 700 m at a baseline length of 16000~m. 

This scheme permits to see the distributions in sufficient resolution and with sufficient counts per bin for all ALMA configurations. And it has the additional useful property that the first bin can exclusively be filled by single-dish (TP) observations and the second bin exclusively by ACA observations. Underperformance in these bins can therefore be used as an indicator that TP and/or ACA data is missing.

A 2D version of the BLD is essentially the familiar "uv coverage" plot which is commonly used in interferometry. The 2D version has the advantage that it also captures the baseline {\it orientation} and thus the elongation and orientation of the synthesized beam ellipse (see the example in Fig. \ref{fig:bldistexamples}d) while the 1D BLD only captures its average diameter. On the other hand, the 1D BLD permits to better visually compare different BLDs (e.g. observed and expected) since one can simply plot one on top of the other.

\subsection{The expected baseline length distribution}
\label{sec:expect}

In order to decide on the array configuration during scheduling and to judge the quality of the observed BLD, we need to derive an expected shape, the Expected Baseline Length Distribution (EBLD) in order to have a reference. This is done by determining the expectation function for the uv coverage from the science goals and filling a histogram with it.

\begin{figure} [htb]
\begin{center}
\begin{tabular}{c} 
\includegraphics[height=6cm]{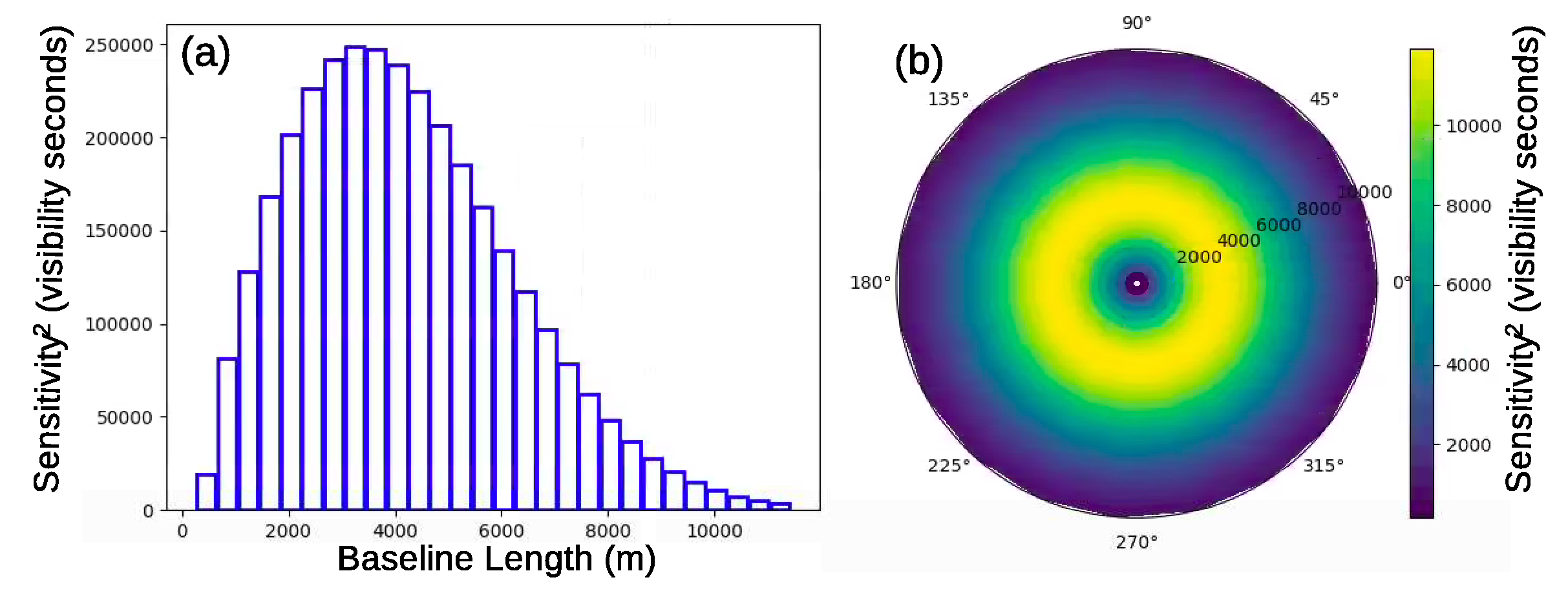}
\end{tabular}
\end{center}
\caption[]
{ \label{fig:expbldexamples} Examples of 1D and 2D analytically calculated expected baseline length distributions for a 30 minute observation with 43 antennas in an extended configuration.}
\end{figure} 

\noindent Here we explore two approaches: 
\begin{description}
\item[a) an analytical approach] which calculates the expected shape from an expected beam shape (derived from the AR) which is Fourier-transformed and then modified by an "inner taper" to account for the LAS requirement.
\item[b) a simulation approach] which
 assumes a perfect interferometer with a maximally filled aperture up to the radius of half the maximum BL length, 
i.e. a disk of the diameter of the maximum BL length (derived from the AR) filled with antennas at a minimum distance (derived from the LAS). 
\end{description}

\subsubsection{Analytical approach}
\label{sec:analytical}
As described above, the science goals of the PI project w.r.t. image properties are captured by the ALMA Observing
Tool (OT) in the proposal submission as the angular resolution (AR) and the largest angular scale (LAS). For an interferometer this
translates into the size of the synthesized beam (FWHM) and the
maximum recoverable scale (MRS) given the angular modes it is
sensitive to, where the latter is defined for ALMA as recovering
10\% of the total flux density of a uniform disk of the
respective size; the primary beam size, i.e. antenna aperture or
mosaic pattern, has to be at least three times larger. As
detailed in the ALMA Technical Handbook \cite{remijan2019},
the scales for a given {\it single} configuration are also
defined via the $5^{\rm th}$ and $80^{\rm th}$ percentile of the
BLD. In our approach we effectively
generalise this latter definition to a match over the whole
BLD and for the case of multi-configuration executions. 

In order to formalise the PI imaging request we assume a Gaussian
target function for the AR for the following reasons:
\begin{enumerate}
\item such aperture is
often adopted in optical applications for reasons of imaging
quality 
\item the functional form is invariant under Fourier
transformation
\item it is the form of the “clean” beam adopted in
the interferometric image reconstruction algorithm CLEAN, i.e.
“dirty” beam = “clean” beam 
\item for the aforementioned reasons,
the ALMA configurations have been designed with a visibility space density optimization
against this functional form for the radial coordinate. 
\end{enumerate}

In our present best version of the analytical BLD creation, the expectation function is
computed as the Fourier transform of the Gaussian beam with
the size given by the AR, and apodized after transformation with
an inner uv taper of an inverted Gaussian (1/r) of the
transformed LAS size. It is  numerically implemented with a regular
rectangular map filling the polar visibility density histogram,
and directly filling the histogram from the analytical expression
considering the pixel sizes. Figure \ref{fig:expbldexamples} shows an example of a BLD computed with this method. Note that for angular scale range
requests less than a factor five, i.e. LAS~$< 5 \times$AR we set
LAS~$:=5 \times$AR, aligned with the baseline distributions of the
ALMA 12-m configurations. 

The expectation function is idealised in terms of the exact radial shape, azimuthal uniformity and overall smoothness. In particular, the designed and actually realised baselines of the antenna array and projection effects during observations inevitably cause deviations, which have to be captured by a metric and evaluated in order to provide guidance for the scheduling task (see Section~\ref{sec:schedPlan}). It is an important point (e.g. references \citenum{briggs1995} and \citenum{boone2013}) that ample sensitivity can be used to improve imaging with a suitable re-weighting of the uv pixels to better resemble the target function and hence improve resolution match, beam circularity and sidelobe suppression (see section \ref{sec:reweight}). 

\subsubsection{Simulation approach}
\label{sec:simulation}
As an alternative to the analytical approach, we are also exploring
the potential benefits of creating the BLD directly by placing antenna positions onto an aperture with uniform spatial density and histogramming their BL lengths.
In order to extract a smooth general shape as in the analytical case, the antennas are placed randomly and the process is repeated a large number of times such that the resulting BLD is an {\it average} of all possible uniform-density array configurations which obey the constraint that all BL lengths are between a given minimum and maximum. The minimum BL length is derived from the LAS, the maximum from the AR. The overall scaling of the BLD is determined by the number of visibilities in the real observation for which the expected BLD is to be derived.

Due to the large-number statistics, the simulation approach naturally results in overall Gaussian-like BLD shapes, quite similar to those of the analytical approach, without assuming this functional form anywhere. See Fig. \ref{fig:bldobsandexp-example} for an example of a BLD derived with the simulation approach.

As the LAS is increased, the BLD of the hypothetical instrument which is constructed in the above process, approaches that of
an equivalent optical telescope with similar image fidelity. 
This way of determining the ideal BLD shape naturally arrives at a sensitivity requirement for very short spacings and thus permits to determine how to include single-dish observations.

\subsection{Comparing the expectation with simulation or observation}
\label{sec:fillfrac}

As described above, a comparison of BLDs with their analytically calculated expectation
is needed both in the context of scheduling and in quality assurance.
To formalise this process, we introduce the "filling fraction" $f$ which is defined for each BLD bin $i$ as
$$
f_i = o_i / e_i \ \ (e_i > 0), f_i = 1 \ \ (e_i = 0)
$$ 
where $o_i$ is the entry in the $i$th bin of the observed BLD while $e_i$ is the corresponding entry in the expected BLD. A filling fraction of 1 means that the observation has fulfilled the expectation. So for bins where $e_i = 0$, we define the filling fraction to be 1.

We also introduce the total filling fraction $f_t = \sum{o_i}/\sum{e_i}$ and the average filling fraction $f_a = \sum{f_i}/n$ where the sums in the expressions for $f_t$ and $f_a$ are computed in the histogram range where $e_i > 0$ and $n$ is the number of bins in that range (which we call {\it expectation range}).

Using these filling fraction definitions, one can construct metrics for the assessment of observed BLDs.
Here, one has to {\it distinguish between science goals for which image fidelity is more relevant and those where it is of minor importance}. If image fidelity is of minor importance (e.g. for detection experiments or observations
of point sources), the resolution at which the observed and expected BLD need to be compared is much reduced.

We have tested a number of possible metrics and arrived so far at a preliminary best choice which is defined as follows:
\begin{enumerate}
    \item Determine observed and expected BLDs with ten equidistant bins over the expectation range.
    \item Compute $f_i$ for each of the ten bins.
    \item If image fidelity is of high importance for the science goal, then, for "pass", require $f_a\geq0.9$ and $f_i > 0.85$ for $i = 0,...,8$ and $f_9\geq1$, i.e. on average the filling fraction should be at least 90\%, at least 85\% in each bin, and at least 100\% in the uppermost bin which decides on the achievement of the AR.
    
    Alternatively, if image fidelity is of minor importance, relax the requirements on bins {0,...,8} and only require $\sum_0^3{o_i}/\sum_0^3{e_i} > 0.85$ and $\sum_4^8{o_i}/\sum_4^8{e_i} > 0.85$, i.e. widen the bins in the lower 90\% of the expectation range such that the shape of the BLD is only roughly measured.
    
\end{enumerate}

\section{SCHEDULING BASED ON BASELINE LENGTH DISTRIBUTIONS}
\label{sec:schedPlan}

\begin{figure} [ht]
\begin{center}
\begin{tabular}{c} 
\includegraphics[height=7.3cm]{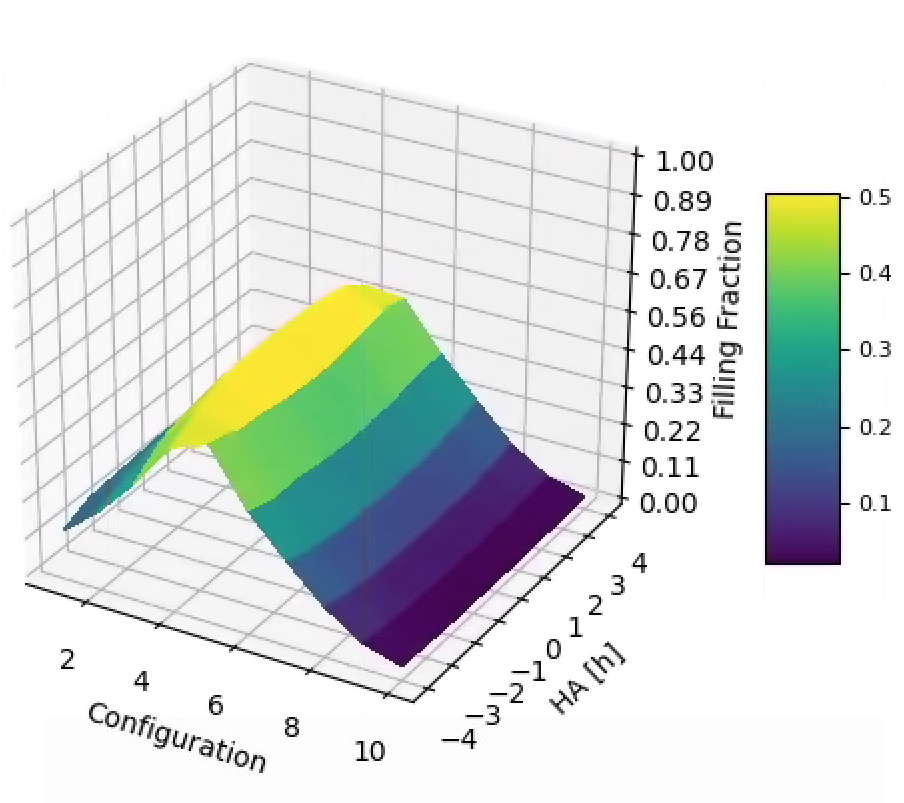}
\includegraphics[height=7.3cm]{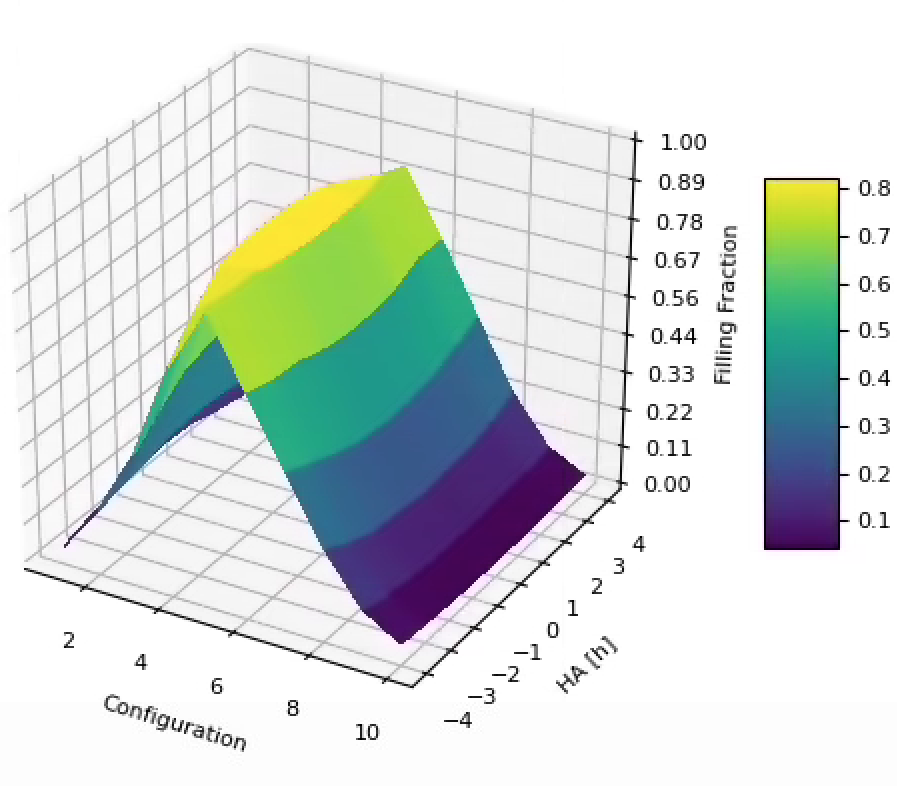}
\end{tabular}
\end{center}
\caption[]
{ \label{fig:SchedFill1} 
An ALMA observation requiring two scheduling block executions in Band 7 with an angular resolution of 0.3 arcseconds can be scheduled optimally in configuration C-4 at hour angles between -2h and +2h, with C-5 also allowed in the second execution. The scheduling options are given as the achieved total filling fractions on the z-axis as a function of configuration on the x-axis and hour angle on the y-axis. This scheduling projection assumes for the second execution that the first execution had been scheduled with maximal total filling fraction. The filling fraction of the second execution refers to the remainder of the first, resulting in over 90\% completion of the observation.}
\end{figure} 

In ALMA long and short term scheduling tasks are distinguished relating to project pressure over seasonal conditions, e.g. weather pattern, day/night phase conditions, configuration schedule, and the daily aspects, e.g. available antenna array, actual weather conditions. In the following we describe our new scheduling approach:

\begin{description}
\item[1 - Determine expected BLD:] Given the science goal parameters of target Declination (DEC), requested sensitivity, LAS, and AR range (i.e. minimum and maximum acceptable AR), the expected BLD of the MOUS is calculated.
\item[2 - Find best-matching ALMA configuration and observation hour angle:] Based on a library of BLDs created from detailed simulations of observations with different ALMA array configurations (see description below, this is different from what is described in section \ref{sec:simulation}), find the optimal ALMA configuration C$_{\rm opt}$ and observation hour angle HA$_{\rm opt}$ which achieve the highest total filling fraction $f_a$.
This is illustrated by Fig. \ref{fig:SchedFill1}.
\item[3 - Observe first/next EB:] When the array is moved into C$_{\rm opt}$ and weather conditions are appropriate, schedule the SB for execution at HA$_{\rm opt}$.
\item[4 - Compare observed BLD with expected BLD:] Using, e.g., the metric described in section \ref{sec:fillfrac}, determine whether the first EB already fulfills the expectation for the MOUS. If it is "pass", declare the observation complete.
When applying the metric, iterate over the whole range of expected BLDs, from the smallest AR value up to the largest.
\item[5 - Determine the remaining necessary observations, i.e. new expected BLD:]
   In case step (4) results in a fail, i.e. the latest EB did not complete the observation, {\it subtract} the BLD of the latest EB from the expected BLD for the MOUS and declare the remainder the new expected BLD.
\item[6 - Re-iterate from step 2:] with the new expected BLD go back to step (2) and determine a new best configuration and HA. Repeat the loop until the observation is declared complete in step (4) or the observing period for the given SB has ended.
\end{description}

\noindent
The library of simulated BLDs needed in step (2) of the algorithm is obtained
by running a simulator for the generation of ALMA visibility data
for a complete range of observatory configurations (presently ALMA has ten nominal configurations numbered 1 to 10), target declinations (DECs in steps of a few degrees), and possible observation hour angles (HAs).
In our study, a prototype of this library was created from a suite of 900 simulations with the CASA\cite{2019arXiv191209437E} task {\tt simobserve} for each of the 10 ALMA configurations on a grid of Declination (in steps of 10 degrees) and hour angle (steps of 1 h above elevations of 30 degrees centered on culmination). 
Figure \ref{fig:SchedFill1} shows the result of the process in step (2) of the above algorithm for a particular example.

An even more accurate scheduling could be achieved by complementing the library of simulated BLDs by simulations of the {\it hybrid} configurations which the array assumes
{\it in between} the nominal configurations (since only few antennas can be moved per day).  During real time scheduling, the actually available antenna array could be considered.

\begin{figure} [htb]
\begin{center}
\includegraphics[height=5cm]{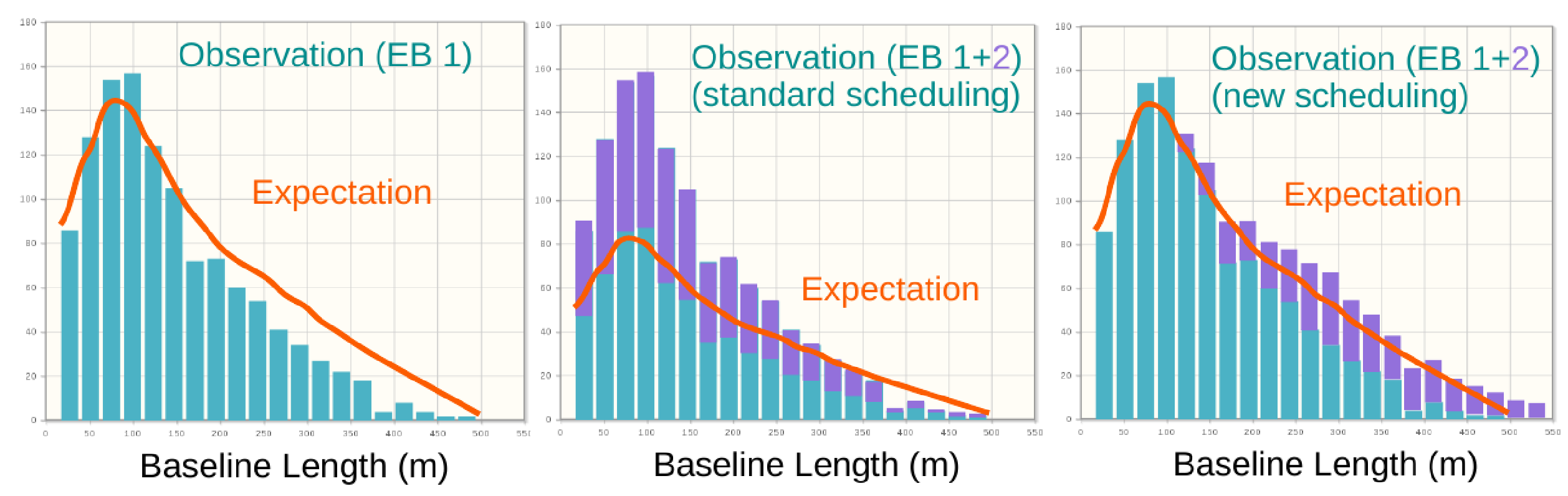}
\end{center}
\caption[]
{ \label{fig:schedadvantage} A generic example of one of the advantages of our new scheduling approach: When a first observation does not fully achieve the expected sensitivity at all baseline lengths, standard scheduling of a repeat-observation would drastically over-observe those baselines which are already well-observed. The new approach determines more accurately the missing observation.
}
\end{figure} 

Figure \ref{fig:schedadvantage} summarizes the main advantage of the iterative scheduling approach w.r.t. observing efficiency.
Note that in reality, the image quality aspects, which are implied by the condition on the BLD, can only be one factor in the scheduling together with other ones imposed by the rules of the observatory:
\begin{itemize}
    \item If the observation of the SB has to be completed by a certain date, e.g. by the end of the observing cycle, this means that the number of available configurations is steadily reduced as the cycle progresses until, near the end, there is only one left. In order to still pass the condition of the BLD metric, it may then be necessary to use a sub-optimal configuration and make up for this by increasing the observation time. This will result in an "over-exposure" at certain
    BL lengths (which can, however, be compensated for by offline reweighting, see section \ref{sec:reweight}).
    \item If one or more EBs have already been observed and the algorithm determines
    a new C$_{\rm opt}$ for the next EB which will only be arrived at much later in the cycle, it may be advisable to give the PI the choice whether to indeed wait or relax the requirements. Alternatively, the observatory can decide, based on quantitative assessment, to
    achieve the completion by investing additional observing time in a sub-optimal configuration.
\end{itemize}

\section{USING BASELINE LENGTH DISTRIBUTIONS IN QUALITY ASSURANCE}

After the BLD-aware scheduling described in the previous section has declared an MOUS fully observed, it is the role of QA in ALMA to verify that the MOUS indeed achieves the science goals. The achieved RMS noise is directly measured in the image
of the representative target in the representative spectral window. This process needs no further improvement.
The achievement of the AR and LAS, however, can for the first time be
assessed properly by comparing the observed BLD ({\it after} the complete flagging and calibration procedure) with the expectation, again using a metric like
the one described in section \ref{sec:fillfrac}: By requiring the filling fraction to be at least 100\% in the uppermost part of the expectation range, we enforce the AR. And by requiring the filling fraction to be close to 100\% in the lowest part of expectation range, we enforce the LAS. Requiring in addition that there be no BLD bins in the expectation range with low filling fraction (e.g. $f_i > 0.85$) ensures that also all intermediate angular scales have been observed with adequate sensitivity.  
For MOUSs where the science goals do not require good image fidelity, we only assess the AR criterion via the upper 10\% of the BLD and treat the rest of the BLD with more coarse binning mostly verifying the overall sensitivity.

\begin{figure} [ht]
\begin{center}
\includegraphics[height=5.5cm]{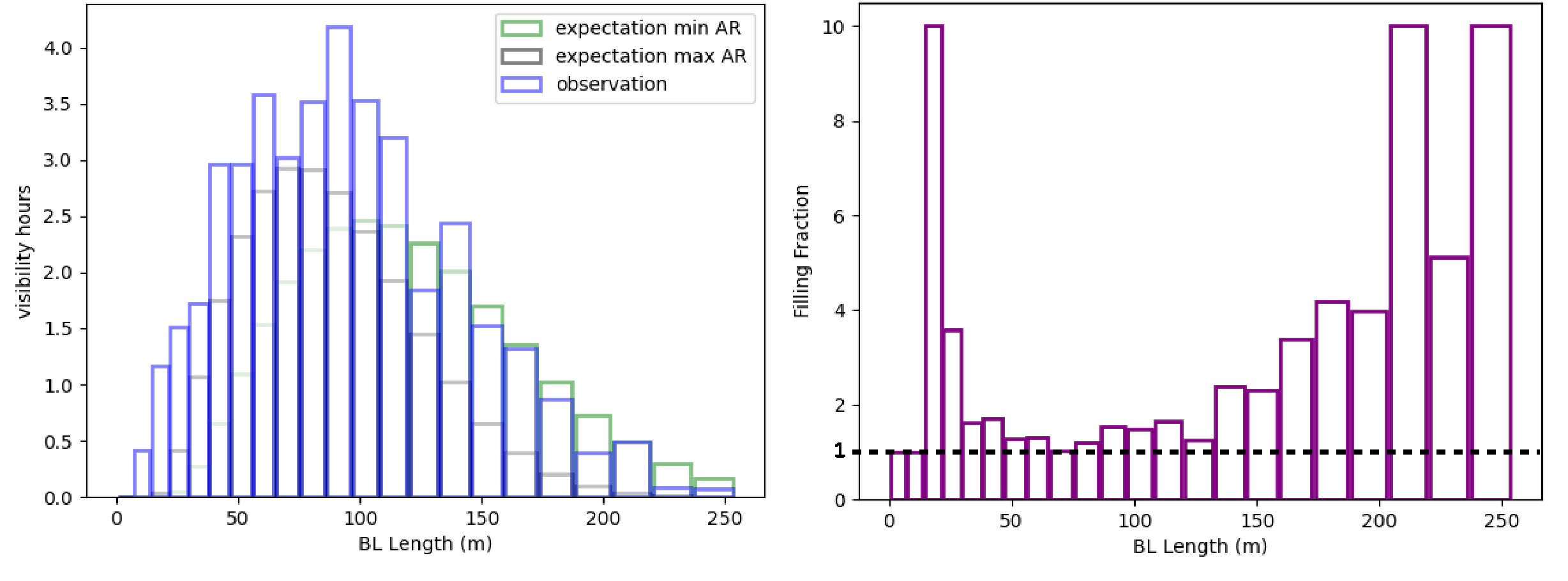}
\end{center}
\caption[]
{\label{fig:fillfrac1example} Example plots from our prototype QA application for an ALMA observation of 19 min duration using 43 antennas (which is equal to the expectation) in a compact configuration. Left: The observed BLD (blue) plotted on top of the two edge cases of the expected BLD, the one for the minimum AR value (2.4 arcsec in this case, green) and the one for the maximum AR value (3.6 arcsec, black). Right: The filling fraction $f$ of this observation w.r.t. the expectation for the maximum AR value. Since $f_i > 1$ in all bins, this observation would pass QA. The high values at both ends show that both the AR and the LAS condition are comfortably fulfilled.
}
\end{figure}

\begin{figure} [ht]
\begin{center}
\includegraphics[height=5.5cm]{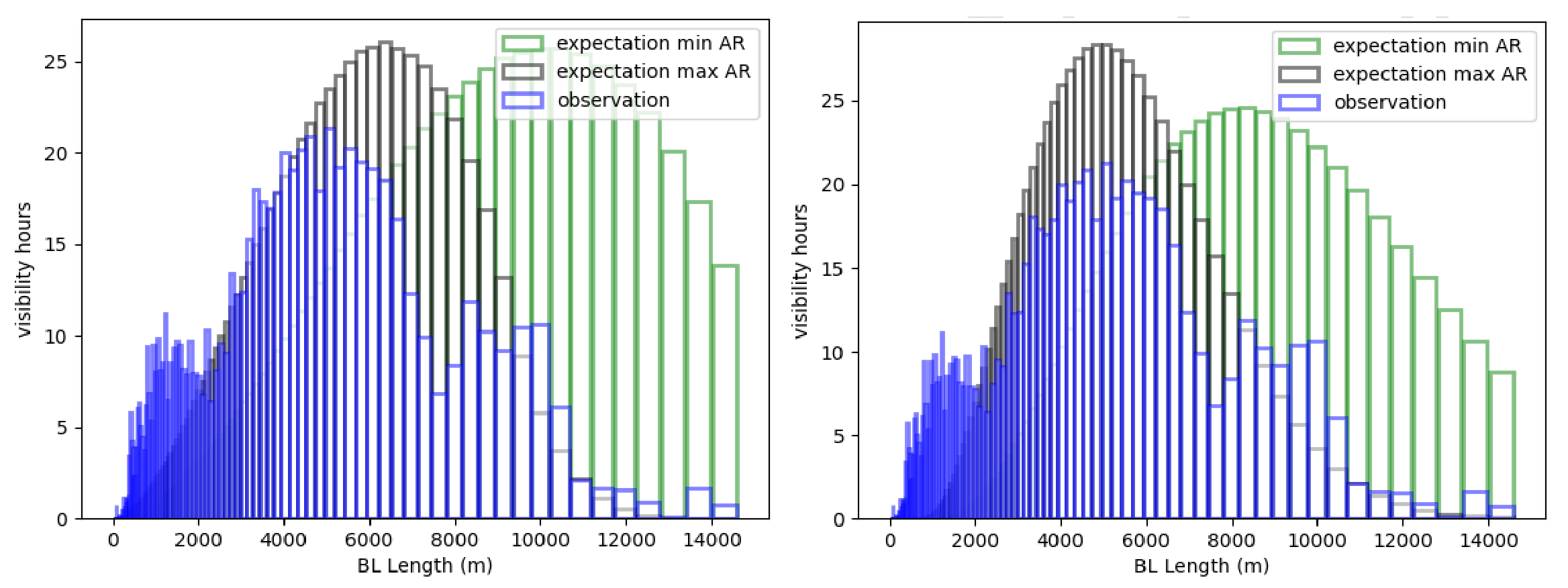}
\end{center}
\caption[]
{\label{fig:bldobsandexp-example}
Another example of plots from our prototype QA application, here for an ALMA observation of 47 minutes duration using 44 antennas in a quite extended configuration at an unusually sub-optimal HA (elevation was around 37$^\circ$ when it could have been around 70$^\circ$ if the observation had been made around HA=0). Left: with the expectation BLDs computed using the simulation approach (section \ref{sec:simulation}). Right: with the expectation BLDs computed using the analytical approach (section \ref{sec:analytical}). The requested AR was between 0.019 arcsec and 0.030 arcsec, the LAS was 0.15 arcsec. In both cases, the AR and LAS requirement is met (for the maximum AR value) while there is certain under-exposure in the intermediate BL length range, which would go unnoticed without inspecting the BLD in this way since the under-exposure is at least partially made up for by the over-exposure at the shortest baselines. 
}
\end{figure}

In our development study we have started to test this approach on several hundred
MOUSs from ALMA Cycles 6 and 7 (see examples of the diagnostic plots in Figures \ref{fig:fillfrac1example} and \ref{fig:bldobsandexp-example}). What we have found so far is that a large fraction of MOUS do pass the new BLD-based condition. In fact, the intermediate to short BL lengths often
seem to be over-observed. This may indicate a possibility to further optimise the ALMA
scheduling. However, we are not yet ready to draw any final conclusions from 
the ongoing study.

In any case, we can already design a new QA2 process for MOUSs which includes the BLD condition:
\begin{enumerate}
    \item After calibration is completed, obtain the observed BLD for the representative target and spectral window from the calibrated data and using the corresponding WEIGHT values.
    \item Compare with the expected MOUS BLD using the agreed metric (e.g. the one described in section \ref{sec:fillfrac})
    \item If the metric results in a "pass", the MOUS has passed the AR and LAS criteria. If it also passes all other QA criteria, it is archived and delivered to the PI.
    \item If the metric results in a fail, i.e. some BL length range is underexposed, we can use the same approach as in scheduling to determine the expected BLD of the {\it missing observation} to complete the MOUS. Essentially, the observed BLD after calibration is subtracted from the original expected BLD of the MOUS, and the difference histogram is the new expected BLD for a re-observation of the SB to be scheduled.
\end{enumerate}

\section{REWEIGHTING VISIBILITIES TO THE IDEAL BASELINE LENGTH DISTRIBUTION SHAPE}
\label{sec:reweight}

It will never be possible to refine scheduling to a point where for each MOUS the observed BLD agrees exactly with the expectation. This is because (a) the limited number of antennas by far does not fill the aperture of the array and (b) considering antenna maintenance and repair tasks, and practicalities during re-location, optimal arrays cannot be provided at all times. In order to ensure a high over-all efficiency
of the observatory, individual MOUSs can only be made to meet the science goals to
a good {\it approximation}.
However, as long as the observatory errs on the side of exceeding the requested
sensitivity, it is possible to correct the {\it shape} of the observed BLD in offline processing.

\begin{figure} [b]
\begin{center}
\includegraphics[height=5.5cm]{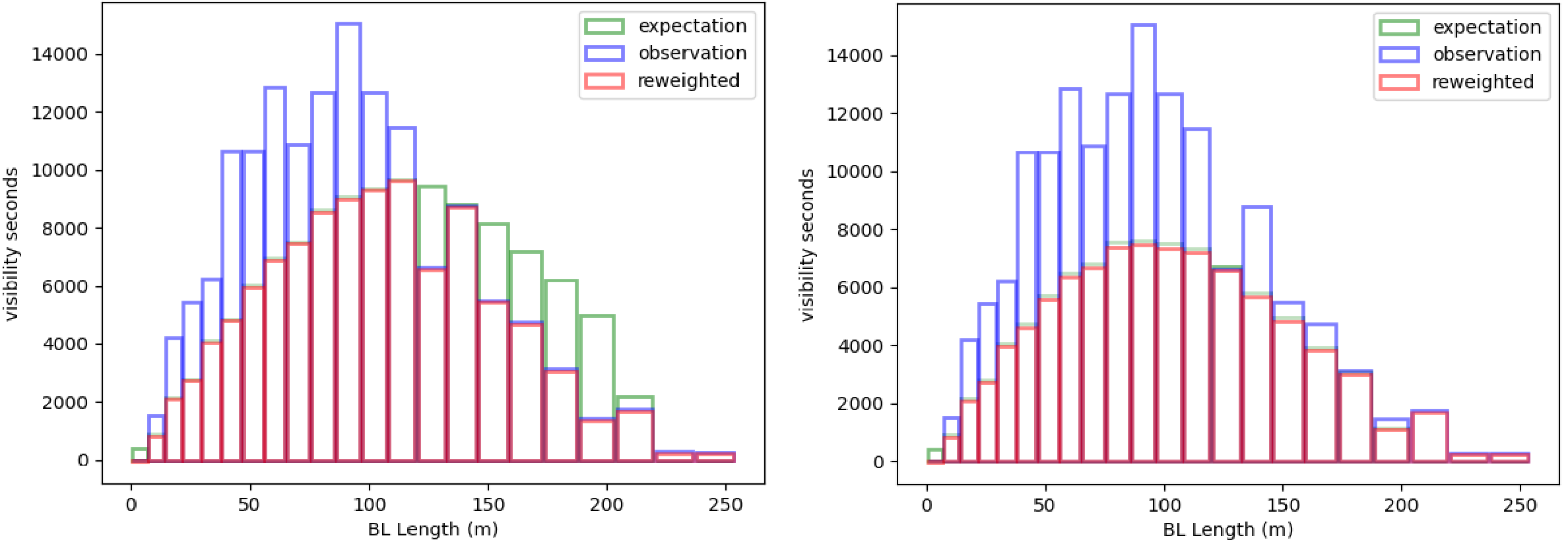}
\end{center}
\caption[]
{ \label{fig:reweightexample} Two examples for the effect of the reweighting of visibilities according to the procedure described in section \ref{sec:reweight}. An observed BLD (blue) is compared to its expectation (green) and for each bin a reweighting factor $\leq 1$ is determined and applied to the weights in the corresponding MeasurementSet. The BLD of the observation after reweighting is shown in red. Left: Here the BLD is partially over-exposed and partially under-exposed. The reweighting brings the formerly overexposed bins down to the level of the expectation while the under-exposed bins are left unchanged. This is a compromise between ideal BLD shape and over-all sensitivity. Right: Here the expectation is well exposed in essentially all bins. The reweighting can achieve a good match with the expectation.}  
\end{figure}

The ideal shape of the observed BLD is the expected BLD as derived according to the methods described in section \ref{sec:expect}. 
To give the final images the optimal fidelity on all accessible angular scales,
the visibilities need to be re-weighted on a per BL length bin basis such that
the reweighted observed BLD has the same shape as the expected BLD. 
Only reweighting factors $< 1$ are permitted (we cannot artificially increase
sensitivity in individual BL length bins). So the reweighting procedure can only
downweight, and thus the improved image fidelity comes at
the price of loss of sensitivity in the downweighted bins and thus only makes
sense where image fidelity is regarded as most important (see also reference \citenum{boone2013}). 

The reweighting factor $w_i$ to be applied for BL length bin $i$ is computed as follows:
$$
w_i = f_{\rm min}/f_i 
$$
where $f_i$ is again the filling fraction and $f_{\rm min}$ is its minimum value across all bins.

As a compromise between optimal sensitivity and optimal image fidelity, one can
modify the above definition of the $w_i$ by only downweighting those bins where $f_i > 1$, i.e. those bins which have been over-observed. Then one has to compute the $w_i$ as
$$
w_i = 1/f_i \ \ \ {\rm for}\ f_i > 1
$$
$$
w_i = 1 \ \ \ {\rm for}\ f_i \leq 1
$$
Figure \ref{fig:reweightexample} shows an example for both cases.

\section{CONCLUSIONS AND OUTLOOK}

The advantages of our new approach to scheduling and QA will mostly come to bear where the relative sensitivity at different angular scales matters: These are mostly deeper exposures on extended objects which aim to properly measure the power spectrum of the image in the range accessible to the instrument. 
While also single-EB observations will get the best possible BLD (and thus best image quality), the real advantage is realised for observations with multiple EBs where we can better ensure that the uv coverage is close to optimal. In particular the extension of our approach to 2D BLDs  will permit QA to address beam ellipticity and holes in visibility space causing beam side lobes. But already working with 1D BLDs will improve image fidelity and minimise over-observation of individual angular scale ranges thereby freeing the instrument for more projects. Figure~\ref{fig:schedadvantage} illustrates the concept.
Standard scheduling which just repeats the same observation until all requirements are met, can be quite wasteful.

Being able to determine more exactly which BL lengths still lack exposure opens the door to a range of scheduling options also including more exotic ideas like the use of {\it sub-arrays}: when, e.g., only exposure in a small range of BL lengths is missing, the observatory may construct an appropriate array from a smaller number of antennas during
engineering time (when many antennas undergo maintenance), and use this array to execute the missing EB
during a time time period which otherwise would not have been available for science observations at all.

Our study has now reached the mid-point and we are turning from the development of concepts, processes, and prototypes to the detailed evaluation of them based on datasets from ALMA archive and simulations. We hope to report our final results at the next SPIE Telescopes and Instrumentation conference after this one.

For ALMA scheduling, as far as we can see now, the new approach would mean a modification in how to select the configuration of the first EB of an MOUS.
For single-EB observations, that would already be all.
Also, for SBs for which only image quality (i.e. low beam sidelobes) is relevant but not fidelity, the changes in scheduling procedure would end here.
Only for multi-EB MOUSs which need good image fidelity at a larger range of scales there would be a change in that we may switch to a different preferred configuration after the first EB(s) have been observed.
Our study still needs to determine how often this would happen and how different the first selected configuration would typically be from the subsequent one.
The point is for these cases not to make scheduling simpler but more efficient, i.e. use less over-observation.
The idea of using sub-arrays is only envisaged for exceptional cases where an MOUS needs a small number of baselines for completion.
Here ALMA could indeed gain overall observing time by using a smaller array during engineering time when conditions permit this.
The proposal is not to split up ALMA into sub-arrays but to use a subset of it while the rest is in maintenance or used for engineering tests.

For the QA process, our approach means that for the first time the LAS condition is explicitly verified and the coverage of intermediate angular scales is assessed at all. The procedure used so far only
verified the overall sensitivity and the sensitivity at the smallest angular scales. 

Furthermore, our approach will permit to conveniently perform QA on {\it groups} of MOUSs (the so-called GOUSs in the ALMA scheduling system already mentioned in section \ref{sec:intro}). For the set of MOUSs in a given GOUS, a single expected BLD can be calculated and thus the combined data from the MOUSs assessed together as one observation.

Finally, the reweighting scheme described in section \ref{sec:reweight} promises to help ALMA users achieve the optimal image fidelity both for individual MOUSs and groups.

As more large radio interferometer arrays start operations around the globe, we note again that our results are generally applicable to all arrays with more than a few antennas.

\acknowledgments 
We would like to thank Carlos de Breuck (ESO) for his support as the contact person for this ESO ALMA internal development study.

\bibliography{almanewmethods} 

\begin{thebibliography}{1}

\bibitem{remijan2019}
Remijan, A., Biggs, A., Cortes, P., Dent, B., Mason, B., Philips, N., Saini,
  K., Stoehr, F., Vila-Vilaro, B., and Villard, E.,  [{\em ALMA Technical
  Handbook Cycle 7}{\nolinebreak\hspace{0.1em}]}, ALMA Doc. 7.3, ver. 1.1
  (2019).
\newblock
  https://almascience.eso.org/documents-and-tools/cycle7/alma-technical-handbook.

\bibitem{nakos2020}
Nakos, T., Francke, H., Nakanishi, K., Petry, D., Stanke, T., Ubach, C.,
  Cerrigone, L., Keller, E., Trejo, A., and Ueda, J., ``{Improving ALMA’s
  data processing efficiency using a holistic approach},'' in [{\em
  Astronomical Telescopes and Instrumentation 2020 (this
  conference)}{\nolinebreak\hspace{0.1em}]},  {\em Proc. SPIE} (2020).

\bibitem{briggs1995}
Briggs, D.,  [{\em High Fidelity Deconvolution of Moderately Resolved
  Sources}{\nolinebreak\hspace{0.1em}]}, New Mexico Institude of Mining and
  Technology (1995).
\newblock PhD thesis.

\bibitem{boone2013}
Boone, F., ``Weighting interferometric data for direct imaging,'' {\em
  Experimental Astronomy}~{\bf 36},  77 (2013).

\bibitem{2019arXiv191209437E}
{Emonts}, B., {Raba}, R., {Moellenbrock}, G., {Castro}, S., {Garcia-Dabo},
  C.~E., {Donovan Meyer}, J., {Ford}, P., {Garwood}, R., {Golap}, K.,
  {Gonzalez}, J., {Kawasaki}, W., {McNichols}, A., {Mehringer}, D., {Miel}, R.,
  {Montesino Pouzols}, F., {Nakazato}, T., {Nishie}, S., {Ott}, J., {Petry},
  D., {Rau}, U., {Reynolds}, C., {Schiebel}, D., {Schweighart}, N., {Steeb},
  J.~W., {Suoranta}, V., {Tsutsumi}, T., {Wells}, A., {Bhatnagar}, S.,
  {Jagannathan}, P., {Masters}, J., and {Wang}, K.~S., ``{The CASA software for
  radio astronomy: status update from ADASS 2019},'' {\em arXiv e-prints} ,
  arXiv:1912.09437 (Dec. 2019).

\end{thebibliography}
\bibliographystyle{spiebib} 

\end{document}